\documentclass[a4paper]{PoS}

\usepackage{graphicx}
\usepackage{float}
\usepackage{subfigure}
\usepackage{csquotes}

\usepackage{xstring}
\usepackage{catchfile}

\usepackage[per-mode=symbol,binary-units]{siunitx}
\DeclareSIUnit\micron{\micro\metre}
\DeclareSIUnit\Gbits{\giga\bit\per\second}
\DeclareSIUnit\mWsqcm{\milli\watt\per\cm\squared}

\usepackage{pgfplotstable}		%% Tabellen einlesen im pfg
\usepackage{tikz}				%% Grafik erstellen		
\usepackage{pgfplots} %% Package laden
\pgfplotsset{compat=1.14}
\usepgfplotslibrary{groupplots}	%% Subplots im pgf
\pgfplotsset{
	colormap = {mycmap1}{
		rgb255=(228,26,28) % Rot
		rgb255=(55,126,184) % Dunkel
		rgb255=(77,175,74) % Grün
		rgb255=(152,78,163) %Pink
		rgb255=(255,127,0) %Braun
		rgb255=(153,153,153) %Dunkelblau
		rgb255=(247,129,191)%Grau
		rgb255=(255,255,51) % Orange
		rgb255=(166,86,40)	%Giftgrün
	}
}
\pgfplotsset{
	colormap = {colorL34}{
		rgb=(1,0,0) % V3
		rgb=(1,0.73,0.06) % V4		rgb=(0.9373, 0.8706, 0) % V4
		rgb=(0.341, 0.733, 0.7843) % G3S
		rgb=(0.141, 0.443, 0.1373) % G34
		rgb=(0.6, 0.6, 0.6) % Global
		rgb=(1, 0, 1) % S3
		rgb=(1, 0, 0.5) % S4
	}
}

\pgfkeys{/pgf/number format/1000 sep={\,}} 	

\tikzset{
	linewidth/.style={semithick},
	line/.style={-, linewidth}, % line style to use for all lines
	tip/.style={->, >=stealth', linewidth}, % tip (arrow)
	rtip/.style={<-, >=stealth', linewidth}, % reverse tip
	bitip/.style={<->, line, >=stealth', linewidth}, % tip in both directions
}

\tikzset{mark size=3.5pt}
\tikzset{line width=0.7pt}
\pgfplotscreateplotcyclelist{mylist}{ %definiert meine eigente cycle list
	every mark/.append style={fill=mycolor1},mark=*, mark options={solid, fill=mycolor1, draw = black}, color=mycolor1, dashed, mark size=3pt\\
	every mark/.append style={fill=mycolor2},mark=square*, mark options={solid, fill = mycolor2, draw = black}, color=mycolor2, dashed, mark size=2.5pt\\
	every mark/.append style={fill=mycolor3},mark=diamond*, mark options={solid, fill = mycolor3, draw = black}, dashed, color=mycolor3\\
	every mark/.append style={fill=mycolor4},mark=triangle*, mark options={solid, fill = mycolor4, draw = black}, dashed, color=mycolor4\\
	every mark/.append style={fill=mycolor5},mark=triangle*, mark options={solid, fill = mycolor5, draw = black, rotate=180}, dashed, color=mycolor5\\
	every mark/.append style={fill=mycolor6},mark=pentagon*, mark options={solid, fill = mycolor6, draw = black}, dashed, color=mycolor6\\
	every mark/.append style={fill=mycolor7},mark=square*, mark options={solid, fill = mycolor7, draw = black, rotate=45}, dashed, color=mycolor7, mark size=2.5pt\\
	every mark/.append style={fill=mycolor8},mark=diamond*, mark options={solid, fill = mycolor8, draw = black, rotate=90}, dashed, color=mycolor8\\	
}

\pgfplotsset{
	linieL34/.style={
		colormap name=colorL34, % activate the defined colormap
		cycle list={[of colormap]},
		every axis plot post/.append style={mark=none, very thick},
		legend style={at={(1,1)},anchor=north east},
	}
}

\pgfplotsset{
	linien/.style={
		colormap name=mycmap1, % activate the defined colormap
		cycle list={[of colormap]},
		every axis plot post/.append style={mark=none, line width=1.2pt},
		legend style={at={(1,1)},anchor=north east},
	}
}

\pgfplotsset{
	messpunkteerr/.style={
		colormap name=mycmap1, % activate the defined colormap
		cycle list={[of colormap]},
		every axis plot post/.append style={mark = none, mark size = 2.5pt, thin, error bars/.cd, y dir = both, x dir = both, y explicit, x explicit, error bar style={line width=0.5pt,solid}},
	}
}

\pgfplotsset{
	messpunktekreiserr/.style={
		colormap name=mycmap1, % activate the defined colormap
		cycle list={[of colormap]},
		every axis plot post/.append style={mark size = 2.5pt, thin, error bars/.cd, y dir = both, x dir = both, y explicit, x explicit, error bar style={line width=0.8pt,solid}},
	}
}

\pgfplotsset{
	beamers/.style={
		colormap name=mycmap1, % activate the defined colormap
		cycle list={[of colormap]},
		every axis plot post/.append style={mark size = 2.5pt, line width=1.5pt, error bars/.cd, y dir = both, x dir = both, y explicit, x explicit, error bar style={line width=0.75pt,solid}},
	}
}

\pgfplotsset{
	messpunkte/.style={
		colormap name=mycmap1, % activate the defined colormap
		cycle list={[of colormap]},
		every axis plot post/.append style={mark = o, mark size = 1.5pt, thick},
	}
}

\pgfplotsset{
	barplot/.style={
		grid = none,
		enlarge x limits = 0.05,
		ymajorgrids=true,
		xmajorgrids=false,
		enlarge y limits = 0,
	}
}

\pgfplotsset{
	compat=1.14, % Keine Ahnung wieso
	cycle list name=mylist, % zB (black white, color, mylist, mark list)
	height= 0.35\textwidth,
	width=0.85\textwidth,
	enlargelimits = false,
	enlarge x limits = 0.03,
	enlarge y limits = 0.03,
	legend cell align={left},
	grid=major,
	scale only axis,
}

\pgfplotsset{ylabsh/.style={every axis y label/.style={at={(0,0.5)}, xshift=#1, rotate=90}}}

\pgfplotsset{compat=1.14,
	emphasize/.code args={#1:#2with#3}{
		\pgfplotsextra{
			\draw[fill=#3] ({axis cs:#1,0} |- {axis description cs:0,0}) 
			rectangle ({axis cs:#2,0} |- {axis description cs:0,1});
		}
	},
	emphasizeopa/.code args={#1:#2with#3and#4}{
		\pgfplotsextra{
			\draw[fill=#3, fill opacity=#4] ({axis cs:#1,0} |- {axis description cs:0,0}) 
			rectangle ({axis cs:#2,0} |- {axis description cs:0,1});
		}
	}
}

\definecolor{mycolor1}{rgb}{1,0,0} % Rot
\definecolor{mycolor2}{rgb}{0,0,1} % Dunkel
\definecolor{mycolor3}{rgb}{0,0.39,0} % Grün
\definecolor{mycolor4}{rgb}{1,0.73,0.06} % Goldenrod
\definecolor{mycolor5}{rgb}{0.93,0.07,0.54} %Pink
\definecolor{mycolor6}{rgb}{0.55,0.27,0} %Braun
\definecolor{mycolor7}{rgb}{0,1,0}%Giftgrün
\definecolor{mycolor8}{rgb}{0.1,0.1,0.1}%Grau
\definecolor{mycolor9}{rgb}{0.6,0.96,1} %Dunkelblau
\definecolor{mycolor0}{rgb}{1,0.69803,0.69803} % Hellrot

\definecolor{color-g3s}{rgb}{0.341,0.733,0.7843}
\definecolor{color-v3}{rgb}{1,0,0} % Dunkel
\definecolor{color-s3}{rgb}{1,0,1} % Dunkel
\definecolor{color-g34}{rgb}{0.141,0.443,0.1373} % Grün
\definecolor{color-v4}{rgb}{0.9373 0.8706, 0} %Orange
\definecolor{color-s4}{rgb}{1,0,0.5} % Dunkel
\definecolor{color-global}{rgb}{0.8157, 0.808, 0.808}%Grau

\pgfplotscreateplotcyclelist{parameter}{ %definiert meine eigente cycle list
	every mark/.append style={fill=mycolor1},mark=*, mark options={solid, fill=mycolor1, draw = black}, dashed, color=mycolor1, thick\\
	every mark/.append style={fill=mycolor3},mark=square*, mark options={solid, fill = mycolor3, draw = black}, dashed, color=mycolor3, thick\\
	every mark/.append style={fill=mycolor6},mark=diamond*, mark options={solid, fill = mycolor6, draw = black}, dashed, color=mycolor6, thick\\
	every mark/.append style={fill=mycolor8},mark=triangle*, mark options={solid, fill = mycolor8, draw = black}, dashed, color=mycolor8, thick\\
}

\pgfplotsset{
	histplot/.style={
		width=\textwidth,
		height=0.3\textwidth,
		enlarge x limits = 0.05,
		legend cell align={left},
	}
}

\title{Mechanics, readout and cooling systems of the Mu3e experiment}

\ShortTitle{Mu3e mechanics, readout and cooling}

\author{\speaker{Frank~Meier~Aeschbacher}$^{a,b}$\thanks{Corresponding author.}, Marin~Deflorin$^c$, and Lars~Olivier~Sebastian~Noehte$^b$, on behalf of the Mu3e collaboration\\
        \llap{$^a$}Paul Scherrer Institut, 5232 Villigen, Switzerland\\
        \llap{$^b$}Universit\"at Heidelberg, Physikalisches Institut, 69120 Heidelberg, Germany\\
        \llap{$^c$}Fachhochschule Nordwestschweiz, Institut f\"ur Thermo- und Fluid-Engineering, 5210 Windisch, Switzerland\\
        E-mail: \email{frank.meier@psi.ch}, \email{noehte@physi.uni-heidelberg.de}, \email{marin.deflorin@fhnw.ch}}

\abstract{Mu3e is an upcoming experiment at Paul Scherrer Institut in the search for the strongly suppressed decay of $\mu\rightarrow eee$. It will use an ultra-lightweight silicon pixel detector using thinned HV-CMOS MAPS chips. Multiple Coulomb scattering is further kept under control with using high density interconnects made of aluminium and operating the detector in a helium atmosphere. More than \SI{1}{\metre\squared} of instrumented surface will produce about \SI{3.3}{\kilo\watt} of heat ($\SI[parse-numbers = false]{\leq 250}{\milli\watt\per\centi\metre\squared}$). Traditional cooling approaches are in conflict with the low-mass requirements, hence a gaseous helium flow cooling system will be implemented. This talk will give a report on the successful data transmission tests with the aluminium interconnects at target speeds of \SI{1.25}{\giga bit\per\second} under realistic condition. The final proof-of-concept of the helium cooling has been achieved with comprehensive cooling simulations and successfully confirmed with laboratory measurements using a full-scale mock-up of the vertex pixel detector.\\
}

\FullConference{The 28th International Workshop on Vertex Detectors - Vertex2019\\
		13-18 October, 2019\\
		Lopud, Croatia}

\begin{document}

% ====================================================================================================
\section{Overview}

\noindent
Reducing the material will remain a major challenge for future vertex detectors. Mu3e, an experiment at PSI to search for the highly suppressed decay $\mu^+ \rightarrow e^+e^-e^+$ \cite{LOI,RP}, has no other option than reducing the detector thickness to the max. The current best limit of $<10^{-12}$ still dates back to 1988 \cite{Bertl19851,Bellgardt:1987du}, obtained by the SINDRUM collaboration using gas detectors. The Mu3e experiment aims to improve this limit by about three orders of magnitude by using state-of-the-art detector technology: monolithic active pixel sensors, helium gas-cooled, thinned down to \SI{50}{\micro\metre} on an ultra-lightweight carrier structure for vertexing and momentum measurements plus SiPM-based scintillating detectors (fibres and tiles) for sub-ns timing. Here, we report on the readout circuits needed to reach this goal and the gaseous helium cooling.

% ====================================================================================================
\section{Mu3e pixel detector concepts and mechanical structure}

\begin{figure}[H]
    \center
    \includegraphics[width=\textwidth]{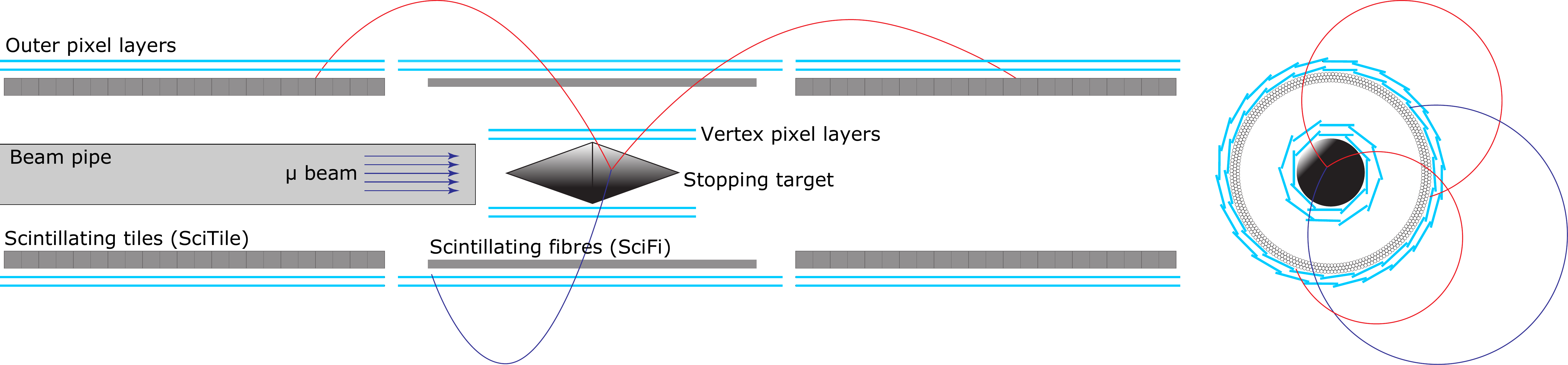}
    \caption{Schematic of the Mu3e experiment with tracks showing a possible decay into $e^+e^-e^+$. Shown in longitudinal (left) and transverse (right) view. Scale indication: Full length about~\SI{1}{\metre}.}
    \label{fig:Mu3eSideView}
\end{figure}

\noindent
The experiment consists of three barrel-shaped detectors, a central one surrounding the muon stopping target and two extending the detector on the upstream and the downstream sides. Each barrel consists of several concentric layers of pixelated silicon detectors ($\approx \SI[parse-numbers = false]{20 \times 20}{\mm\squared}$ active area per chip) and scintillating detectors, where the central barrel surrounds a thin plastic target (hollow double-cone) where muons are stopped, awaiting their decay. The detector is placed inside a $\approx\SI{2.8}{\meter}$ long superconducting solenoid with a field of \SI{1}{\tesla} and an inner bore diameter of \SI{1}{\meter}.
In this configuration, decay tracks with a maximum momentum of \SI{53}{\MeV} follow a helical trajectory passing through the central detectors (two layers of vertex pixels\footnote{Named after their main function to provide vertex information of the decay happening on the target. These are the innermost layers and exist in the central barrel only.} $\rightarrow$ one layer of scintillating fibre $\rightarrow$ two layers of outer pixel\footnote{Named after their location, providing momentum information through reconstruction of recurling tracks. These are the outermost layers, same design in all three barrels.}. Depending on the azimuthal angle of the particle's trajectory, the particle is either curling in the central barrel or stopping in the scintillating tile detector of either the upstream or downstream barrel. An illustration with a possible decay is shown in Fig.\ref{fig:Mu3eSideView}. More details can be found elsewhere \cite{arXiv:1610.02021}.

%\begin{figure}[H]
%    \center
%    \includegraphics[width=\textwidth]{./graphics/general/Mu3eWithTracks3.jpg}
%    \caption{Rendering of the Mu3e experiment with tracks showing a possible decay into $e^+e^-e^+$. }
%    \label{fig:Mu3eWithTracks3}
%\end{figure}

% How much do we really want to say about the mechanics? In the talk, I've just shown the standard slides and I didn't put any emphasis beyond just what is needed to motivate the rest of the talk.
% We could e.g. reference 
% * P. Cattaneo and A. Schöning MEG II and Mu3e status and plan(link is external), EPJ Web of Conferences 212, 01004, 2019.
% * arXiv:1802.09851
% which are more recent publications showing some aspects of the detectors.
% And we should cite N. Berger et al. Ultra-low material pixel layers for the Mu3e experiment(link is external), arXiv:1610.02021 (physics.ins-det), JINST 11 C12006 (2016)., which is sort-of the predecessor of this

% ====================================================================================================
\clearpage
\section{Ultra-thin aluminium high density interconnects for sensor readout}
\begin{figure}[H]
    \center
    \includegraphics[width=1\linewidth]{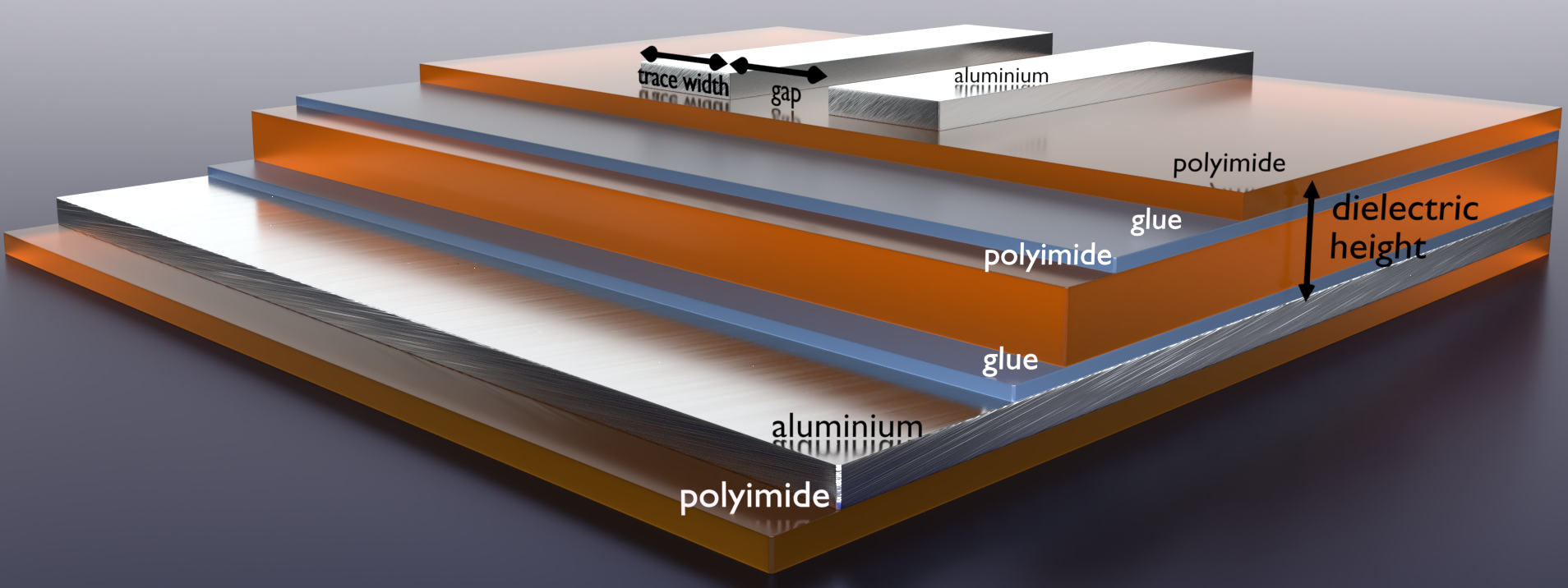}
	\caption{HDI layer stack from bottom to top: \SI{10}{\micro \meter} polyimide, \SI{12.5}{\micro \meter} aluminium, \SI{5}{\micro \meter} glue, \SI{25}{\micro \meter} polyimide, \SI{5}{\micro \meter} glue, \SI{10}{\micro \meter} polyimide, \SI{12.5}{\micro \meter} aluminium \cite{MaNoehte}.}
    \label{fig:layerstack}
\end{figure}

\noindent
The fraction of radiation lengths, $X/X_0$, generated by the detector ladders comes from the sensor and the high density interconnects~(HDI). 
A value of $X/X_0 \approx 0.115\%$ can be achieved by thinning the silicon sensors to \SI{50}{\micron} and using a thin aluminium-based HDI. The latter consists of two layers of a laminate made of \SI{10}{\micron} polyimide and \SI{12}{\micron} aluminium each. The two laminates are glued together with epoxy and feature a polyimide spacer of \SI{25}{\micron} in between, see Fig.~\ref{fig:layerstack}. 
All data and clock connections are made as differential pairs routed in the upper layer where the lower layer provides a solid reference plane from either ground or power lines. 
High-speed data lines have a maximum length of \SI{18}{\cm} and are operated at \SI{1.25}{\Gbits}, calling for tight impedance control.

\begin{figure}[htb]
    \center
    \includegraphics[width=1\linewidth]{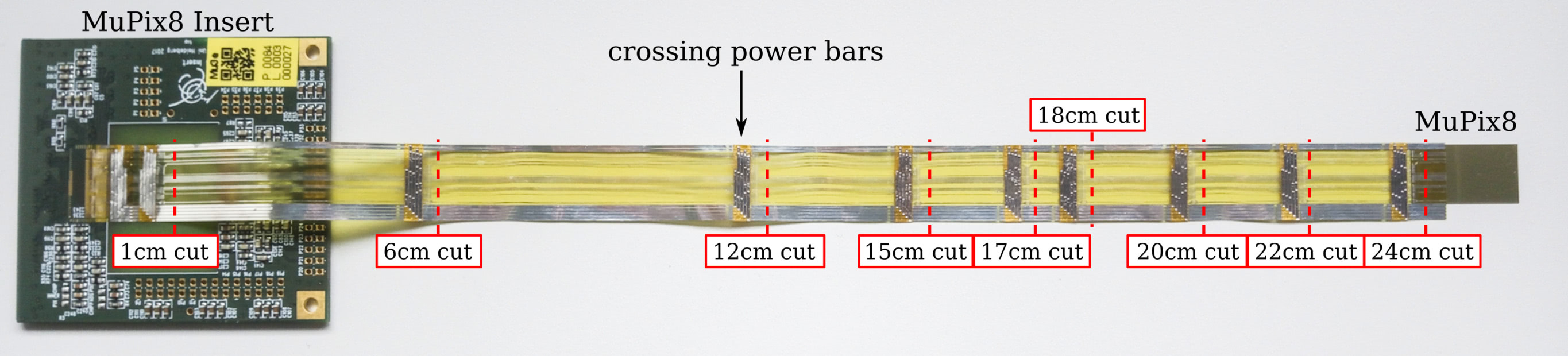}
	\caption{Demonstrator HDI shown with cutting points, bonded to the MuPix8 and the MuPix8 Insert \cite{MaKroeger}}
    \label{pict:HDI_with_cuts}
\end{figure}

\begin{figure}[htb]
    \centering
    \hfill
	\subfigure[HDI bonded to the MuPix8 sensor.]{\label{pict:Bond_chip}\includegraphics[width=0.4\linewidth]{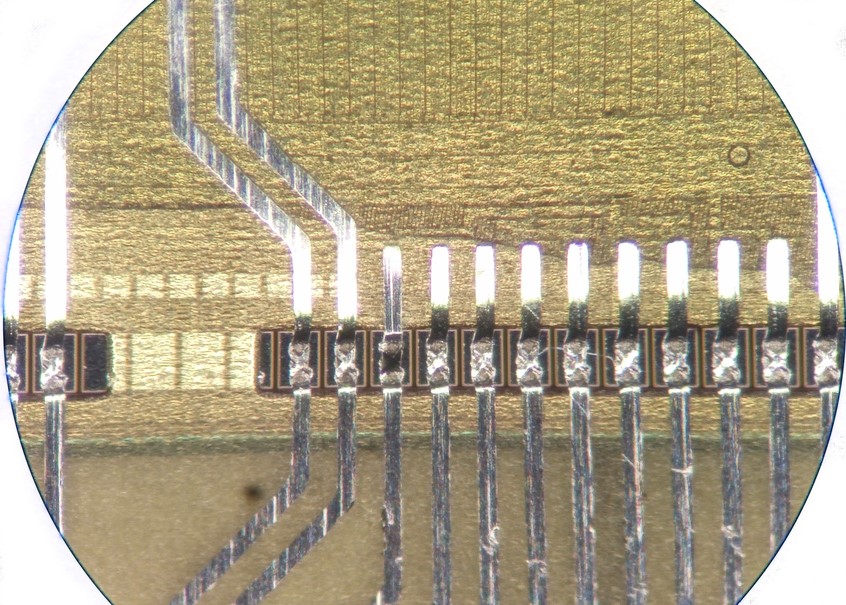}}\hfill
    \subfigure[SpTAB close-up]{\label{pict:Bond_insert}\includegraphics[width=0.4\linewidth]{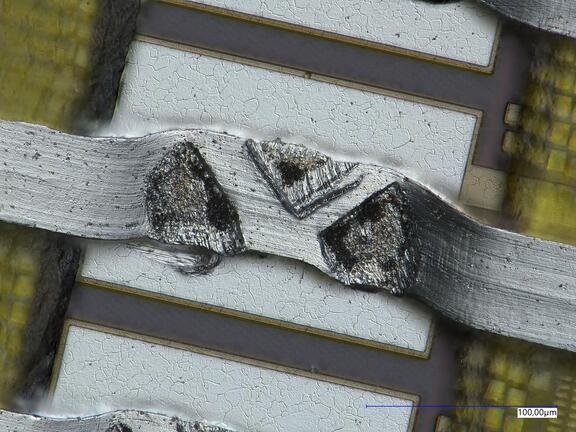}}\hfill
    \caption[Pictures of SpTAB]{Demonstrator HDI bonded via SpTAB}\label{pict:bonds}
\end{figure}

A demonstrator HDI has been designed for a test under realistic conditions, connecting the MuPix prototype MuPix8\footnote{MuPix is a high voltage monolithic active pixel sensor for the pixel detector of the Mu3e experiment. MuPix8 was the latest nearly full size prototype at the time of this study.}~\cite{mupix8_2018,mupix8_2019} to the Mu3e standard lab readout system \cite{MaNoehte}.
The demonstrator has a signal length of \SI{24}{\cm}, which is longer than the \SI{18}{\cm} needed in the Mu3e experiment. It is designed to be trimmed down to shorter lengths if needed, see Fig.~\ref{pict:HDI_with_cuts}.
The HDI was manufactured by the Ukraine company LTU.
%The HDI connects on the one end to a MuPix prototype, the MuPix8, and on the other side to a printed circuit board.
%The MuPix8 was chosen, because at the time of writing it is the only MuPix sensor that can be bonded to the HDI.
The technology used for bonding is known as \emph{Single point Tape Automated Bonding} (SpTAB), which is superior to ball-grid- or wire-bonding due to its reliability and material efficiency.
The bond is established by pressing the desired trace of the HDI through an opening in the polyimide down onto the bond-pad of the sensor.
Through pressure and ultrasonic vibration the metal of the trace and the pad fuse.
A microscope picture of SpTAB can be seen in Fig.~\ref{pict:bonds}.

\begin{figure}[H]
    \centering
    \hfill
    \subfigure[FPGA reference]{\label{pict:reference_eye}\includegraphics[width=0.5\linewidth]{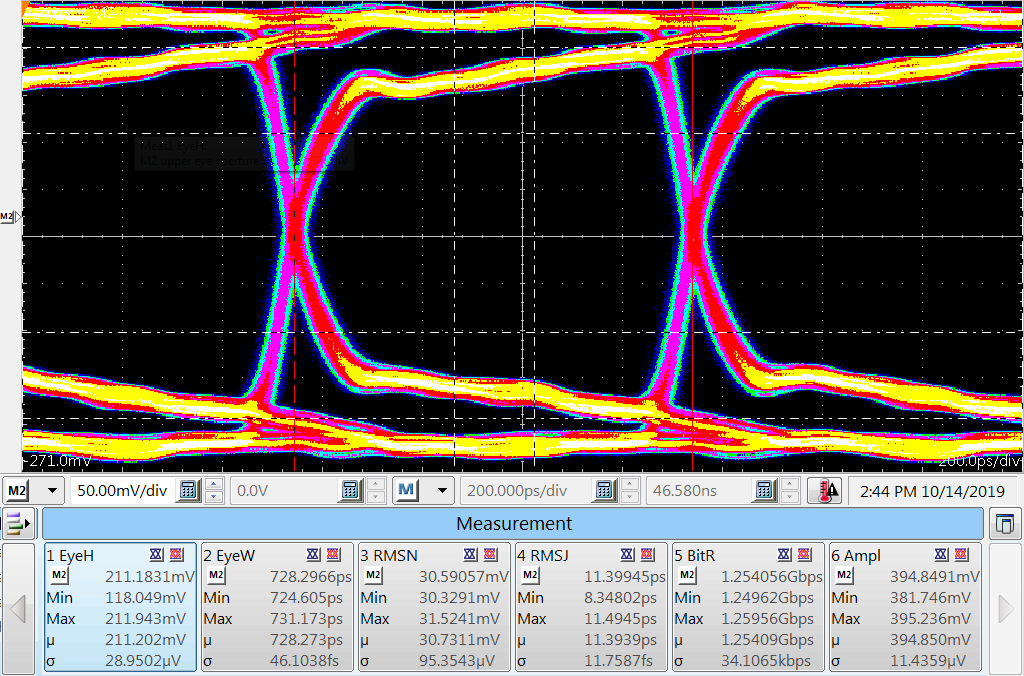}}\hfill
    \subfigure[HDI differential microstrip]{\label{pict:microstrip_eye}\includegraphics[width=0.5\linewidth]{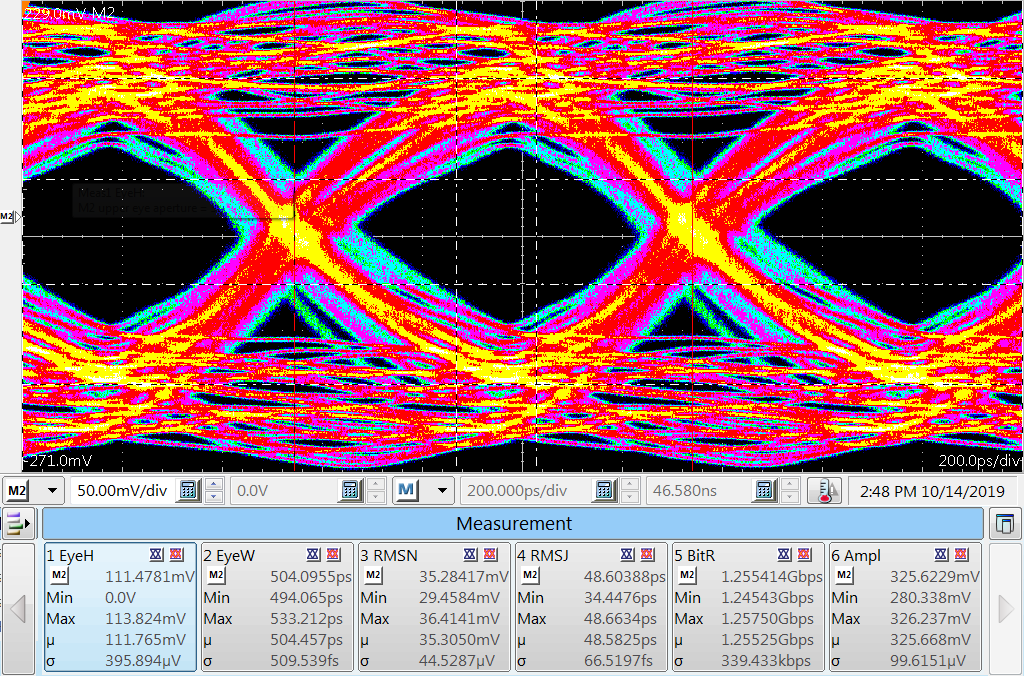}}\hfill
    \caption{Eye diagram measurement}\label{pict:Eyes}
\end{figure}

\begin{figure}[htb]
    \centering
    \hfill
    \subfigure[HDI differential microstrip time domain reflectometry measurement]{\label{pict:TDR}\includegraphics[width=0.5\linewidth]{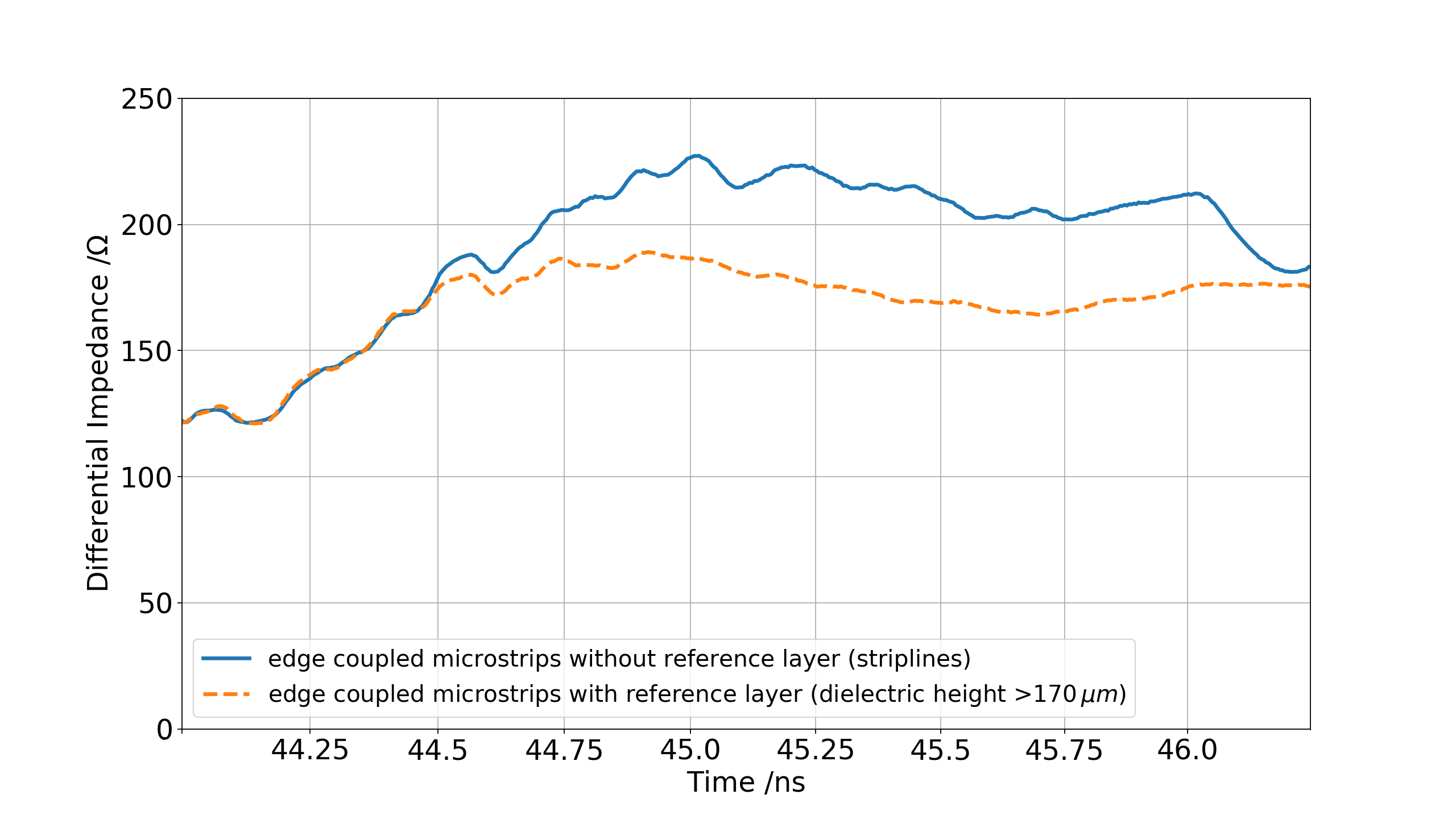}}\hfill
    \subfigure[atlc2 simulation compared to analytic model as functions of the dielectric height]{\label{pict:simulation}\includegraphics[width=0.5\linewidth]{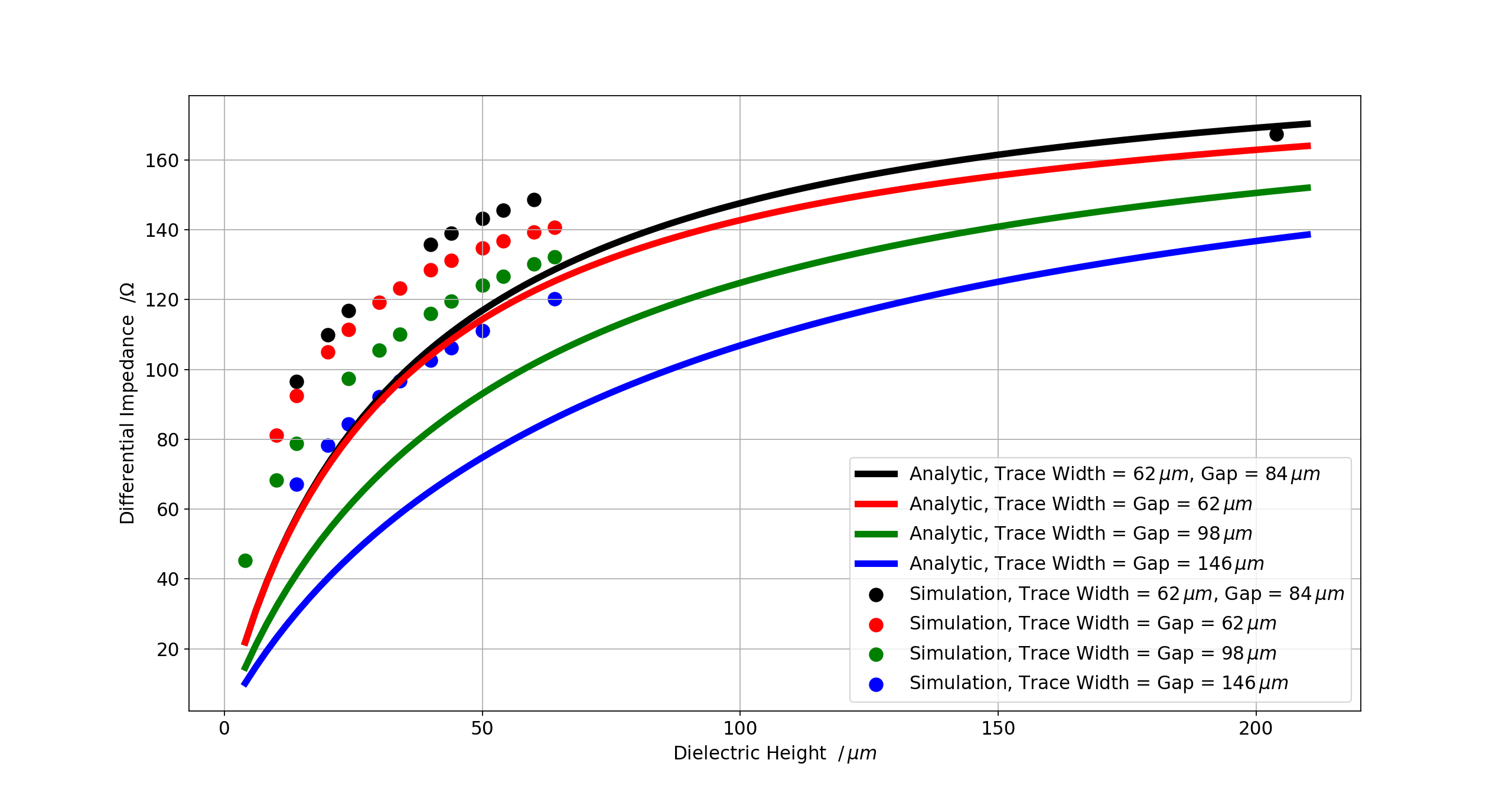}}\hfill
    \caption{Microstrip impedance measurements, simulations and calculations}\label{pict:Impedance}
\end{figure}

\begin{figure}[hbt]
    \center
    \includegraphics[width=0.5\linewidth]{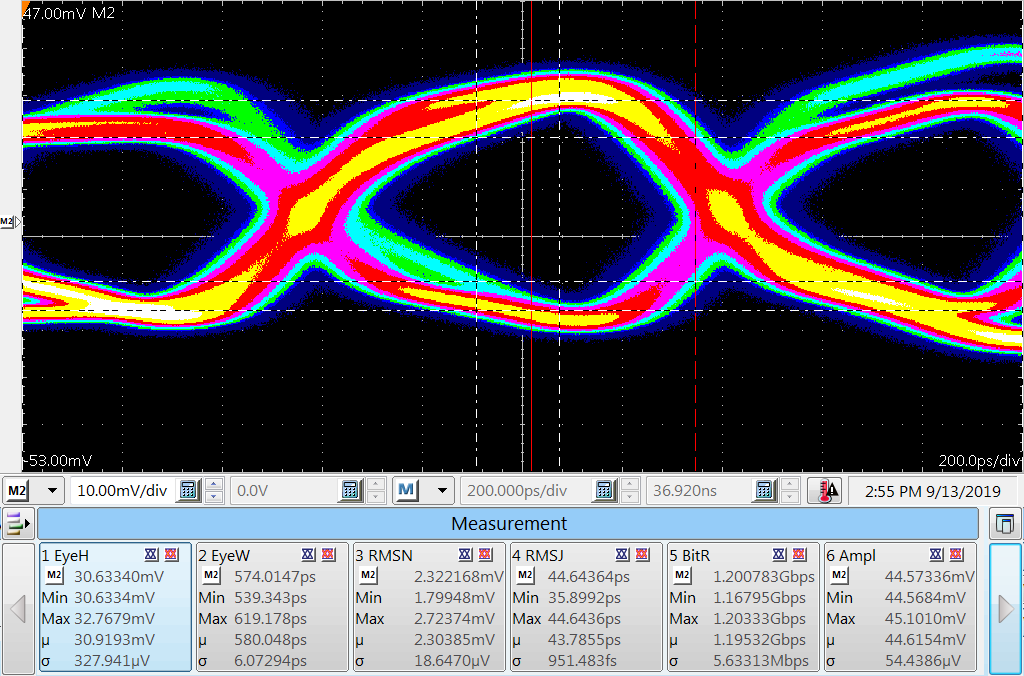}
	\caption{1.25\,Gb/s MuPix8 DataOut4 signal transmitted over a \SI{24}{\centi\meter} long HDI. The high-pass filters is attached via the an additional PCB. EyeH~=~eye height, EyeW~=~eye width, RMSN~=~noise standard deviation, RMSJ~=~jitter standard deviation, BitR~=~measured bit rate, Ampl~=~amplitude}
    \label{pict:MuPix8_Flex_eye}
\end{figure}

%\begin{figure}[htb]
%    \center
%    \includegraphics[width=0.8\linewidth]{./graphics/hdi/SpTAB_micro.jpg}
%	\caption{SpTAB HDI trace to MuPix8 bond pad}
%    \label{pict:SpTAB_micro}
%\end{figure}
The differential traces on this HDI were tested by using a pseudo random bit sequence (PRBS7) for a bit error rate test.
The bit error rate was determined to be less than \num{2e-15}~(95\% CL).
For further analysis, eye diagrams have been measured.
The eye diagram in Fig.~\ref{pict:reference_eye} shows the reference signal, i.e.~the device under test is replaced with a female to female SMA connector.
Fig.~\ref{pict:microstrip_eye} shows the eye diagram with the differential transmission line on the HDI as device under test.
The eye height is determined to be \SI{112}{\milli\volt}.
Compared to the reference from Fig.~\ref{pict:reference_eye} this results in an effective 5.5\,dB loss in signal strength.
The jitter shows a data dependent increase to \SI{49}{\pico\second} caused by the slew-rate in combination with the variety of amplitudes from different bit patterns.
Capping the amplitude to be just larger than the eye height would reduce the jitter.
The eye diagrams as well as the bit error rates have been measured before adding the bottom layer, essentially making the differential pairs \emph{edge coupled striplines} with the expected higher impedance compared to \emph{edge coupled microstrips}.
The resulting impedance distribution, shown in the time domain reflectometry measurement in Fig.~\ref{pict:TDR}, does not match the LVDS impedances standard.
Completing the intended layer stack in post production shows the expected effect of lowering the differential impedance, if a reference layer is present (see orange, dashed curve in Fig.~\ref{pict:TDR}).
For the achieved glue thickness, the impedance measurement is compatible with the simulations using \cite{atlc2} and calculations from \cite{Hammerstad1, Hammerstad2} shown in Fig.~\ref{pict:simulation}.
LTU is capable of decreasing the glue thickness to approximately \SI{5}{\micron}.
Even lower impedances at LVDS level are possible, as shown in~\cite{BaNoehte}.
%Since impedance matching is the key to a reflection-less signal transmission, a time domain reflectometry measurement was done, see Fig.~\ref{pict:TDR}.
%The solid blue curve shows the results for the test specimen without a ground layer as reference layer/potential, essentially representing an \emph{edge coupled stripline pair} with the expected higher impedance.
%In order to achieve a comparable layer stack to the intended one, a reference layer has been added post production. 
%The glue thickness achieved was between \SI{120}{\micro\meter} and \SI{170}{\micro\meter}, much larger than anticipated. 
%With more sophisticated procedures, glue thicknesses down to the required \SI{5}{\micro\meter} are possible but were not accessible for this test.
%Nevertheless, the added reference layer shows the expected effect, lowering the impedance of the differential pair.
%The impedance profile of the tested specimen clearly does not match the \SI{100}{\ohm} LVDS standard, which is no surprise for a stack with dimensions as shown in Fig.~\ref{fig:layerstack}.
%Lower impedances at LVDS level are possible, as shown e.g.~in \cite{BaNoehte}.
%\begin{figure}[htb]
%    \center
%    \includegraphics[width=0.8\linewidth]{./graphics/hdi/HDI_TDR_measurement_w_reflayer.png}
%	\caption{HDI differential microstrip time domain reflectometry measurement: maximum~=~\SI{240}{\ohm}, minimum~=~\SI{124}{\ohm}, mean~=~\SI{197}{\ohm}}
%    \label{pict:TDR}
%\end{figure}
%The impedance has been calculated \cite{Hammerstad1, Hammerstad2} and simulated \cite{atlc2} for different configurations of trace width, gap, and dielectric height (Fig.~\ref{fig:layerstack}).
%The resulting differential microstrip impedances are pictured in Fig.~\ref{pict:simulation}.
%The uncertainty can be assumed to be in the order of 5\% according to \cite{Hammerstad1, Hammerstad2} and \cite{atlc2}.
%At a dielectric height of \SI{204}{\micro\meter}, the simulated and analytically calculated results are compatible with the measurement from \ref{pict:TDR}.
%\begin{figure}[htb]
%    \center
%    \includegraphics[width=0.8\linewidth]{./graphics/hdi/Zdiff_against_dielectric_height.png}
%	\caption{atlc2 simulation compared to analytic model as functions of the dielectric height}
%    \label{pict:simulation}
%\end{figure}
The ultimate test for this HDI is the already mentioned demonstrator equipped with a MuPix8, which exceeds the requirement of \SI{18}{\centi\meter} readout length.
The sensor has been operated at the required readout speed of 1.25\,Gb/s over the full length of \SI{24}{\centi\meter} available on the HDI.
Additional RC-filters are added through a printed circuit board, bonded close to the sensor.
The filters stabilise the voltage controlled oscillator and thus increase the readout stability.
Since pull-up resistors are absent on the MuPix8, the signal is very weak.
The eye diagram in Fig.~\ref{pict:MuPix8_Flex_eye} shows the eye height of the data output from the MuPix8 over the HDI to additional two printed circuit boards required for operating the MuPix8.
%It measures an eye height of \SI{31}{\milli\volt}.
%The Mupix8 can drive the signal stronger but apparently without pull-up resistors not faster.
%To mitigate intermediate voltage levels for single bits, the signal driver is operated at maximum output power and the signal amplitude is limited using a de-emphasis. 
%This limitation leads on the one hand to defined voltage levels for ones and zeros but on the other hand it leads to a low swing.
%Nevertheless the amplitude is still strong enough for the receiver, leading to the successful readout at 1.25\,Gb/s.

\clearpage
% ====================================================================================================
\section{Pixel cooling with gaseous helium}

\noindent
Gaseous helium at ambient pressure and temperatures above \SI{0}{\celsius} has been chosen as the cooling fluid. In helium, compared to air, multiple Coulomb scattering is much lower and the technically achievable heat transfer is higher. Computational fluid dynamics simulations have been carried out to guide the technical design. Two heat-load scenarios have been pursued for best estimation of cooling needs. The realistic scenario uses a heat density of \SI{250}{\mWsqcm}, motivated by measured heat dissipation values with recent MuPix chips under realistic conditions and an estimate of power losses in the HDI. The pessimistic scenario uses \SI{400}{\mWsqcm}, representing the upper limit of what is expected to be cooled away in a technically feasible manner. Within uncertainties, simulations showed a linear relationship between the two scenarios. This allowed to save CPU time by simulating only the pessimistic scenario to full detail and only those results are shown here.

\begin{figure}[H]
    \center
    \def\svgwidth{\textwidth}
    {\scriptsize 
        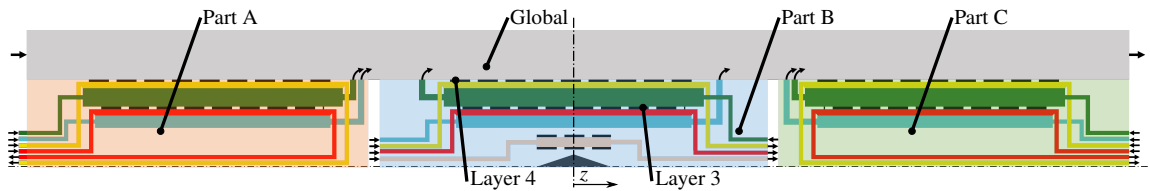
    }
    \caption{Scheme of the cooling flow and parts. For better visibility, the vertical scale has been enlarged and the distance between barrels is exaggerated. The outer layers of all three parts are identical and comprise four helium circuits each. The central station includes the vertex detector with one dedicated helium circuit. The global flow surrounding all barrels forms the vent of about half the circuits.
    }
    \label{fig:L34ABC_schematic}
\end{figure}

The concept foresees 13 different coolant circuits (1~for the vertex detector, $3\times 4$~for each outer layer station, see Fig.~\ref{fig:L34ABC_schematic}), totalling to a helium mass flow of about \SI{55}{\gram\per\second}. As an indication, the expected $\Delta T$ between the inlet and outlet is then about \SI{18}{\kelvin}.\footnote{The pixel detector consists of 2844~chips (108 in the vertex detector, 912 in one of the three outer layer stations), giving about \SI{1.14}{\metre\squared} of active instrumented surface ($\SI[parse-numbers = false]{20\times 20}{\milli\metre\squared}$ active area per chip, neglecting the chip periphery) or about \SI{1.3}{\metre\squared} including chip peripheries. The pessimistic (optimistic) scenario leads to about \SI{5.2}{\kilo\watt} (\SI{3.3}{\kilo\watt}) of dissipated heat. The specific heat capacity of gaseous helium used is \SI{5.2}{\kilo\joule\per\kg\per\kelvin}.}
A cooling plant, situated close to the detector, will provide the required helium flow. The simulated pressure drops across the detector require compression ratios of about 1.1, allowing the use of miniature turbo compressors as an energy-efficient option for generating the helium flow. A pilot plant for engineering purposes is currently under development and will be tested with the vertex detector mockup described below.

The simulation studies have been performed using Ansys\textsuperscript{\textregistered}~CFX,~Release~18.2. A full report is available elsewhere \cite{MaDeflorin}.

\subsection{Vertex detector cooling}
\begin{figure}[hbt]
    \center
    \subfigure[Vertex detector cut view]{ {\footnotesize 
    	\label{fig:HeL12cutview} 
    	\def\svgwidth{0.45\textwidth} 
    	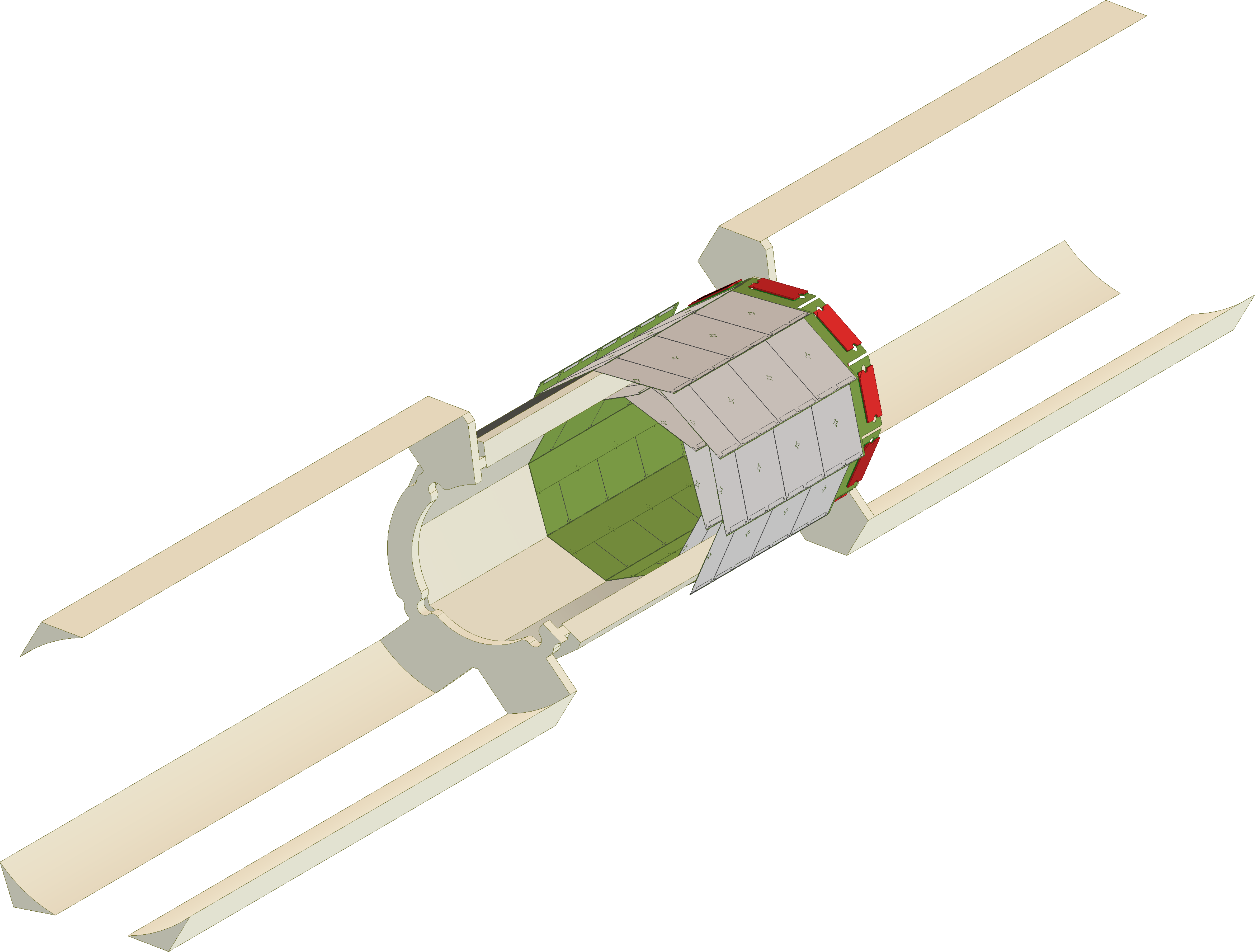
   	}}
    \subfigure[Cross-section view of vertex detector]{
    	\label{fig:L12Xsec}
    	\def\svgwidth{0.36\textwidth}
%% Accompanies image file 'L12Xsec.pdf' (pdf, eps, ps)
\begingroup%
  \makeatletter%
  \providecommand\color[2][]{%
    \errmessage{(Inkscape) Color is used for the text in Inkscape, but the package 'color.sty' is not loaded}%
    \renewcommand\color[2][]{}%
  }%
  \providecommand\transparent[1]{%
    \errmessage{(Inkscape) Transparency is used (non-zero) for the text in Inkscape, but the package 'transparent.sty' is not loaded}%
    \renewcommand\transparent[1]{}%
  }%
  \providecommand\rotatebox[2]{#2}%
  \newcommand*\fsize{\dimexpr\f@size pt\relax}%
  \newcommand*\lineheight[1]{\fontsize{\fsize}{#1\fsize}\selectfont}%
  \ifx\svgwidth\undefined%
    \setlength{\unitlength}{216.67218018bp}%
    \ifx\svgscale\undefined%
      \relax%
    \else%
      \setlength{\unitlength}{\unitlength * \real{\svgscale}}%
    \fi%
  \else%
    \setlength{\unitlength}{\svgwidth}%
  \fi%
  \global\let\svgwidth\undefined%
  \global\let\svgscale\undefined%
  \makeatother%
  \begin{picture}(1,0.98259588)%
    \lineheight{1}%
    \setlength\tabcolsep{0pt}%
    \put(0,0){\includegraphics[width=\unitlength,page=1]{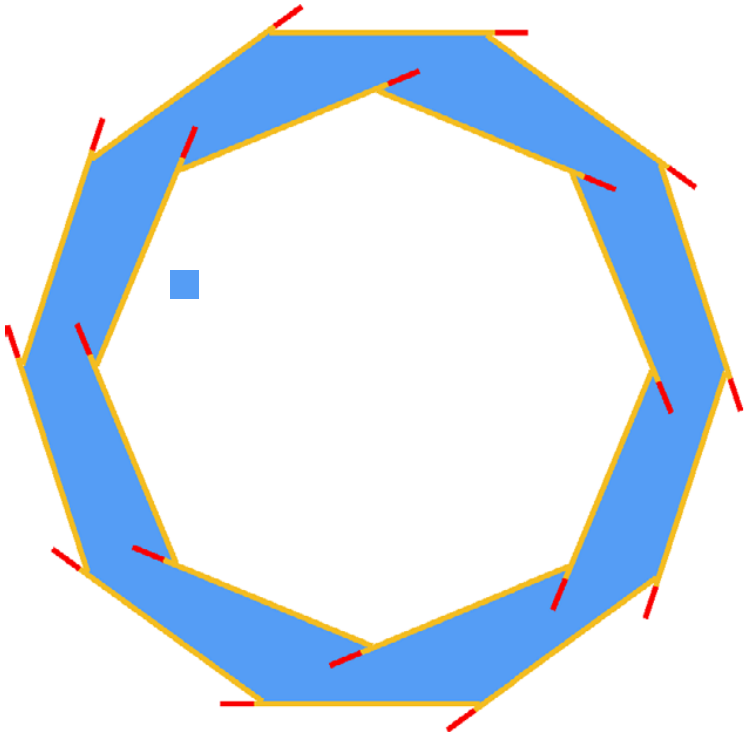}}%
    \put(0.276634,0.5878256){\color[rgb]{0,0,0}\makebox(0,0)[lt]{\lineheight{1.25}\smash{\begin{tabular}[t]{l}Gaseous helium flow\end{tabular}}}}%
    \put(0,0){\includegraphics[width=\unitlength,page=2]{./graphics/cooling/L12Xsec.pdf}}%
    \put(0.27854916,0.49908447){\color[rgb]{0,0,0}\makebox(0,0)[lt]{\lineheight{1.25}\smash{\begin{tabular}[t]{l}Active matrix\end{tabular}}}}%
    \put(0.276634,0.41310913){\color[rgb]{0,0,0}\makebox(0,0)[lt]{\lineheight{1.25}\smash{\begin{tabular}[t]{l}Chip periphery\end{tabular}}}}%
  \end{picture}%
\endgroup%

    }
    \caption{Vertex detector cooling concept. (a) shows the detector with the inlet and outlet ducts. Some structures removed for enhanced visibility. The number of inlet and outlet ducts in the real setup is four. The blue volume in the cross-sectional view (b) between the two barrels forms the coolant volume. The pixel chips are modelled as individual silicon volumes with different heat dissipation densities where red denotes the \emph{periphery} with a higher power density than the \emph{active pixel matrix} shown in yellow.}
    \label{fig:vertexHeCooling}
\end{figure}

\noindent
The vertex pixel detector is the one closest to the stopping target and it is cooled down by one helium circuit. It consists of two \SI{12}{\cm} long concentric barrels with in total 108~chips, shown in Fig.~\ref{fig:vertexHeCooling}. Roughly \SI{200}{\watt} of heat will be dissipated by this detector in the pessimistic scenario. Simulations have been performed with a mass flow of \SI{2}{\gram\per\second}, a value that can be provided given the space restrictions with acceptable pressure drops in the ducts.

\begin{figure}[hbt]
    \center
    \subfigure[Tape heater. Shown are the back and front sides.]{\label{fig:L12tapeheater}\includegraphics[width=.427\textwidth]{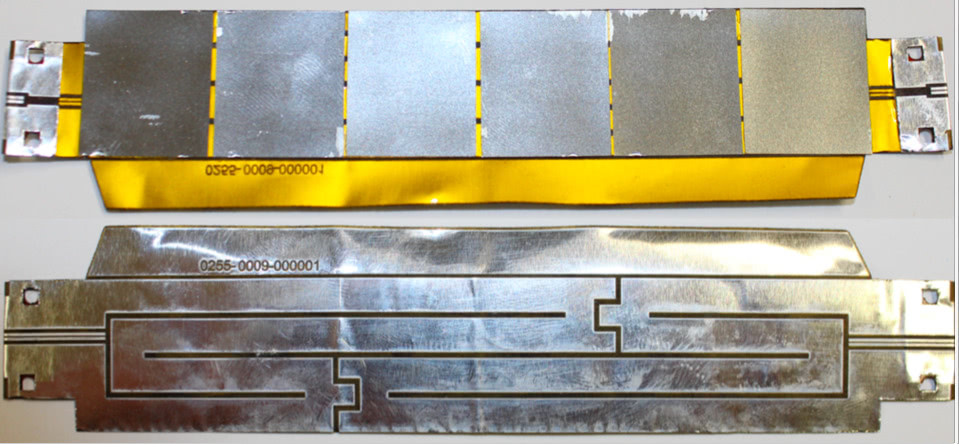}}
    \subfigure[Fully assembled mock-up]{{\footnotesize 
    	\label{fig:L12CoolingFoto}
    	\def\svgwidth{0.473\textwidth}
    	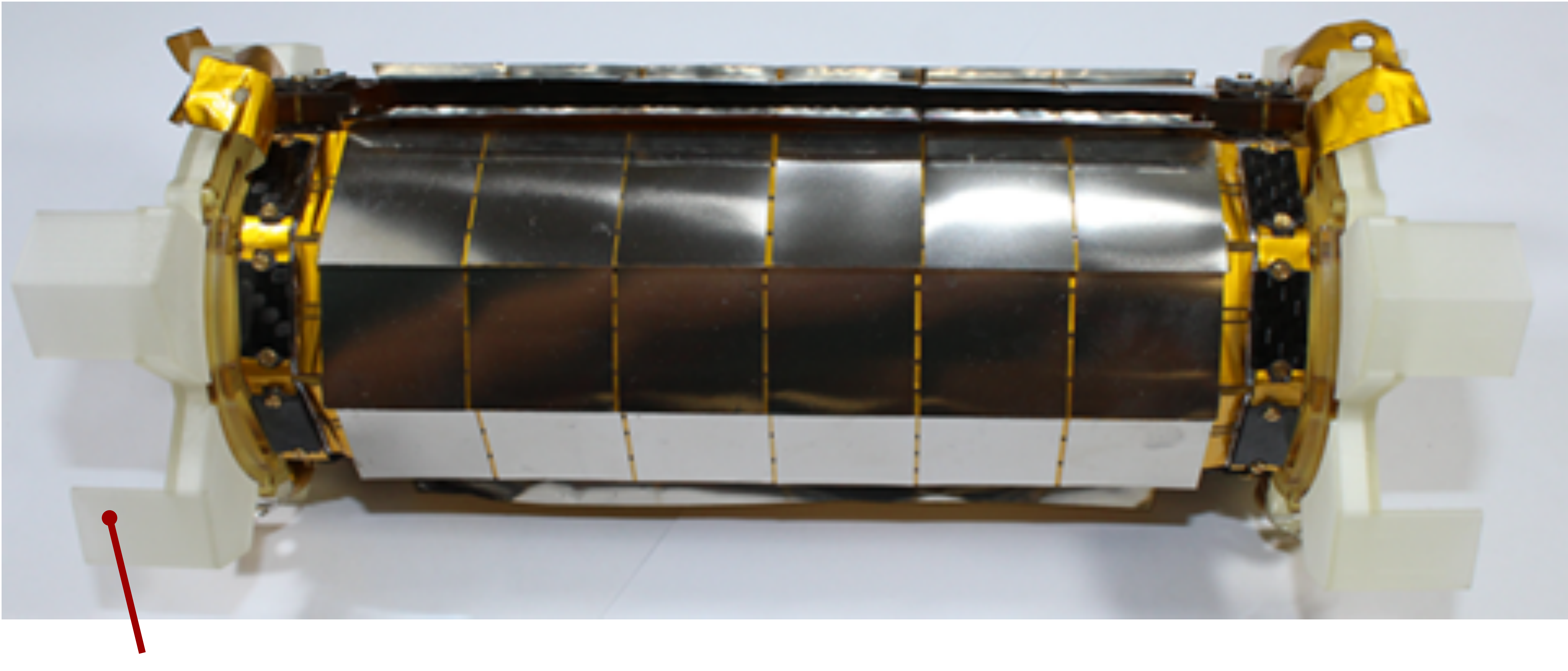
   	}}
    \caption{Thermo-mechanical vertex detector mock-up. Dimensions, material thicknesses and gas flow match current detector design.}
    \label{fig:tapeheatermockup}
\end{figure}

\paragraph{Mock-up.}
To validate the simulations, a full-sized thermo-mechanical mock-up has been built, shown in Fig.~\ref{fig:tapeheatermockup}. The material is an aluminium-polyimide laminate, each component \SI{25}{\micron} thick. Laser-structuring allowed controlled evaporation of the aluminium to form suitable meanders with an approximate resistance of \SI{110}{\milli\ohm}. One ladder consist of two such meanders powered from the short edges. Stainless steel plates with a thickness of \SI{50}{\micro\metre} have been glued as dummy silicon pixel chips. The thickness and the heat capacity of such a ladder approximately match the expected values for a detector ladder. 18 such ladders are mounted using milled plastic structures out of polyetherimide. All ladders are then connected in series through connecting pieces of the same laminate and cables. Connection to the ladders are made using Samtec ZA8H interposers with $7 \times 12$ contact positions, equally split among the two poles. The overall resistance of this object is roughly \SI{4}{\ohm}, enabling to study various power scenarios with voltages and currents easily provided by standard laboratory power sources. This mock-up matches the detector design in terms of mechanical dimensions, fluid path, support structure and achievable heat density.

\begin{figure}[hbt]
	\center
	\subfigure[Simulated temperature on the outer layer of the mock-up.]{%
            \label{fig:L12_T_CFD}
            \includegraphics[width=.427\textwidth]{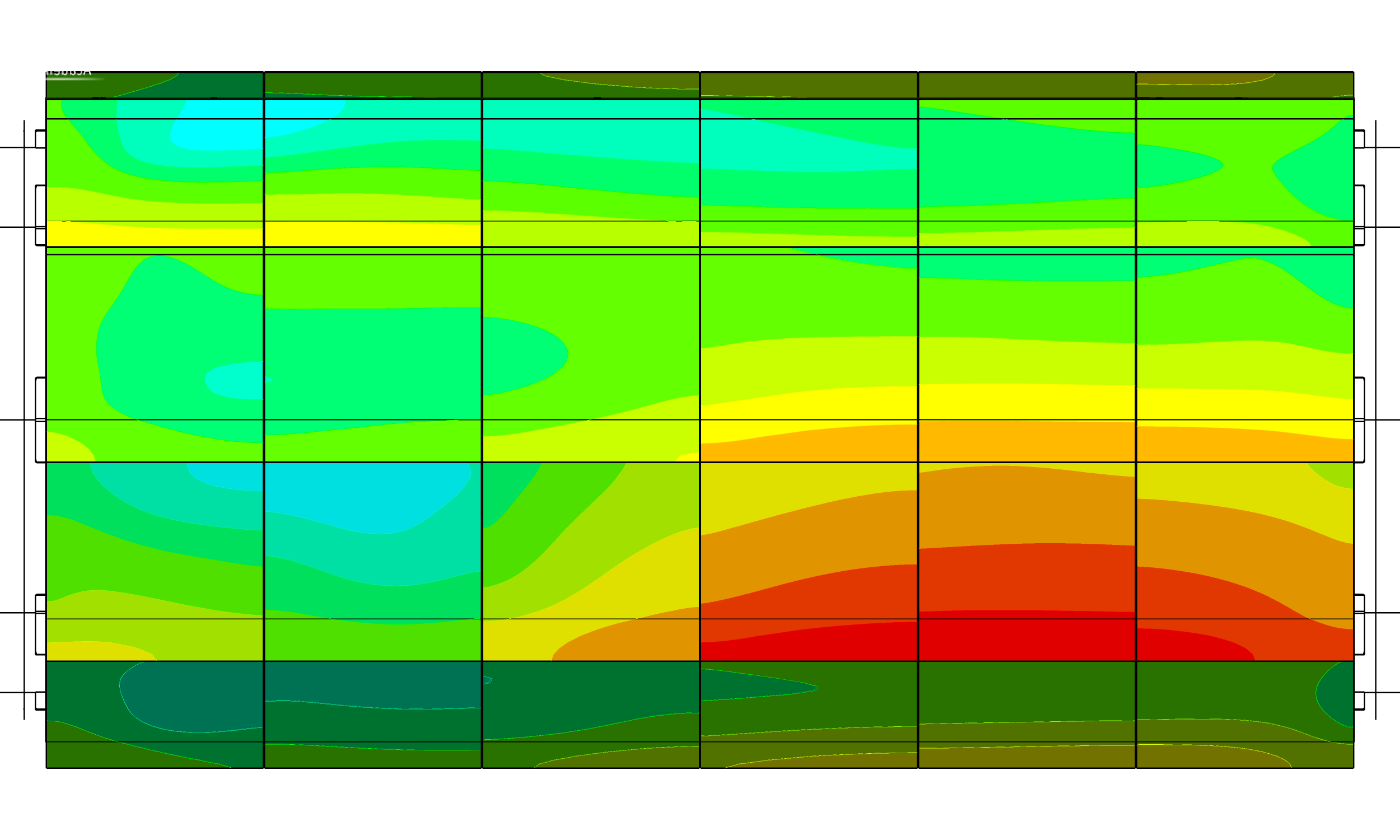}
            \includegraphics[width=.05\textwidth]{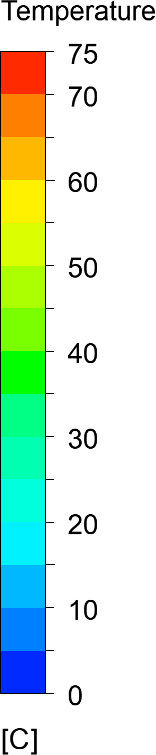}
        }
	\subfigure[Measured temperature on the outer layer of the mock-up using an infrared camera.]{\label{fig:L12_T_Meas}\includegraphics[width=.58\textwidth]{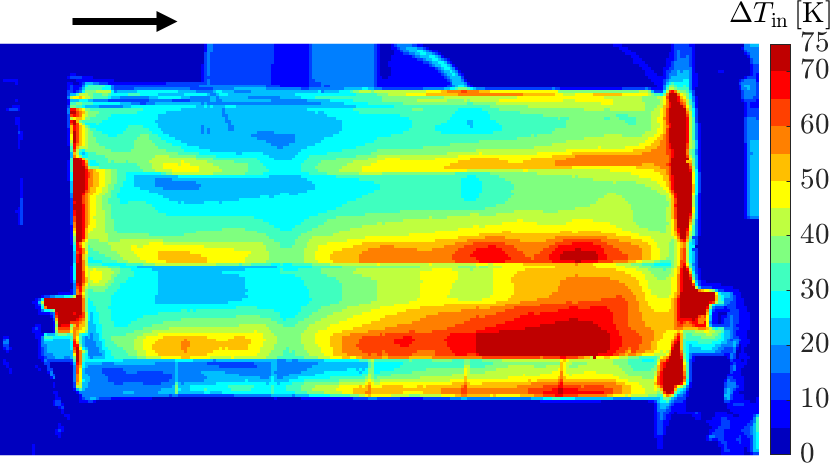}}
	\caption{Temperature obtained by measurement and CFD-simulation. Angle of view of simulation has been carefully matched to the camera view.}
	\label{fig:L12_Temp}
\end{figure}

\paragraph{Results.}
An example simulation result is shown in Fig.~\ref{fig:L12_T_CFD}. An intriguing hot area (lower right) lead to further investigations. The choices for placing the coolant inlets was limited because space is also needed for the electrical connections to and from the detectors, hence inlets were not placed keeping constant arc-lengths in between. In addition, the overlaps from the non-active chip periphery zone (Fig.~\ref{fig:L12Xsec}) break the symmetry in the flow chamber. Both factors give raise to a vortex with local back-flow.

The study using the mock-up showed very good agreement with predictions from simulations, as can bee seen in Fig.~\ref{fig:L12_T_Meas}. Coolant was helium at \SI{2}{\gram\per\second} from a compressed gas bottle. The formation of the vortex has been verified, enhancing the confidence towards the simulation models. The hot zones at the very end stem from non-optimal electrical connections outside the active volume using the aluminium laminate. In the detector, this would be copper connections and therefore do not point to any issues here.

\subsection{Outer layer detector cooling}

\begin{figure}[hbt]
    \center
    \includegraphics[width=0.6\textwidth]{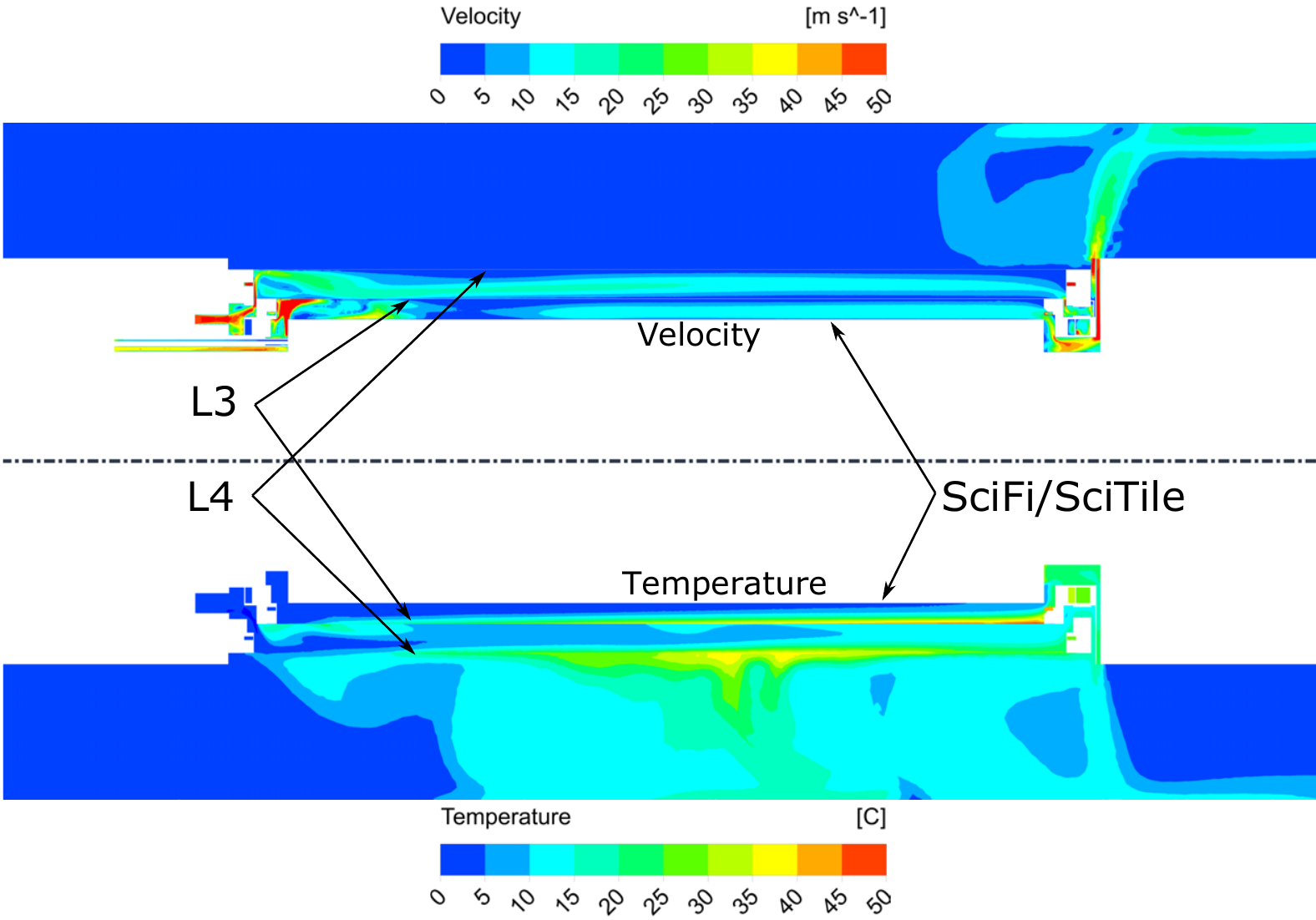}
    \caption{Results of the full simulation. Shown are the coolant velocities (top) and the temperatures (bottom) in a cut view. The inner boundaries are given by the scintillating detectors (SciFi or SciTile). Observe the increased velocities in the global flow outside the detector due to the hot gas vents.}
    \label{fig:L34_flow_temp}
\end{figure}

The outer layers are much larger than the vertex detector. So far, only simulation studies of the full detector have been performed. As the three barrels are identical, an example of a simulation for one barrel is shown in Fig.~\ref{fig:L34_flow_temp} (simulations with all three barrels have been performed and show similar results). Those simulations clearly show the feasibility of helium cooling. Of importance is the temperature distribution at the inner boundary, defined by the scintillating detector inside the barrel. The structures there are especially thermally sensitive (scintillating tiles and fibre ribbons have silicon photomultipliers in their vicinity requiring low and stable temperatures) and thermal influx to them should be minimised. The simulations suggest a very low thermal influx, grace to the almost laminar flow in the annulus formed by the scintillating detector and the third pixel layer (inner layer of an outer station). A full thermo-mechanical barrel is currently under construction and will be tested under real-world conditions, equipped with sensors for temperature, pressure and relative humidity.

% ====================================================================================================
\section{Conclusions}
The Mu3e experiment is at a point where it is time to build the first detector modules.
Reaching the goal of operating a MuPix sensor over an ultra-thin high density interconnect at the full readout speed of 1.25\,Gb/s is a great achievement.
Moreover, the HDI exceeds with \SI{24}{\centi\meter} the length requirements and convinces with a bit error rate of \num{2e-15}~(95\% CL). This even under aggravated circumstances as additional pull-up resistors on the sensor as well as a completed layer stack for impedance matching will improve the transmission quality.

The feasibility of the helium cooling has been shown with extensive simulation studies. Very good agreement to laboratory measurements has been found using a mock-up setup for a vertex detector. This comprises only a subset of the detector, accessible for study with bottled helium. Confidence in the concept has been gained to continue with the design of the cooling plant followed by further studies.

% ====================================================================================================
\section*{Acknowledgements}
The aluminium HDIs have been manufactured by \emph{Research and Production Enterprise LTU LLC}, Kharkiv, Ukraine. We gratefully thank for the always superb support and the high quality of their product.

We gratefully thank the technicians and engineers involved in the technical design work and manufacturing of prototypes. Without this tedious work such a detector could not be designed and tested. The workshops did a marvellous job. We thank Thomas Rudzki for the skilful assembly of the tape heater detector mock-up.

% ====================================================================================================

%{\tiny 
%    This revision: \texttt{\commitshort} on branch \texttt{\branch}.
%}

\begin{thebibliography}{99}
    \bibitem{LOI}
        %A.~Blondel, S.~Bravar, M.~Pohl, S.~Bachmann, N.~Berger, A.~Sch\"oning, D.~Wiedner, P.~Fischer, I.~Peric, M.~Hildebrandt, P.-R. Kettle, A.~Papa, S.~Ritt, G.~Dissertori, Ch. Grab, R.~Wallny, P.~Robmann and U.~Straumann,
        A.~{Blondel} et~al.,
        \textit{``Letter of intent for an experiment to search for the decay $\mu \rightarrow eee$''}, 2012.
        Available online at \texttt{https://www.psi.ch/de/mu3e/documents}.

    \bibitem{RP}
        A.~{Blondel} et~al.,
        {\em ``{Research Proposal for an Experiment to Search for
        the Decay $\mu \rightarrow eee$}''}, arXiv e-prints, January 2013,
        arXiv:1301.6113 (physics.ins-det).

    \bibitem{Bertl19851}
        W.~Bertl et~al., [SINDRUM Collaboration], {\em ``Search for the decay $\mu^+ \rightarrow e^+e^+e^-$''},
        Nucl. P, B 260(1) 1 -- 31, 1985.

    \bibitem{Bellgardt:1987du}
        U.~Bellgardt et~al., [SINDRUM Collaboration],
        {\em ``{Search for the Decay $\mu^+ \rightarrow e^+ e^+ e^-$}''}, Nucl.Phys., B299 1, 1988.

    \bibitem{arXiv:1610.02021}
        N.~Berger et~al. {\em ``Ultra-low material pixel layers for the Mu3e experiment''}, arXiv:1610.02021 (physics.ins-det), JINST 11 C12006 (2016).

    \bibitem{mupix8_2018}
        H.~Augustin et~al. {\em ``MuPix8 – Large Area Monolithic HVCMOS Pixel Detector for the Mu3e Experiment''}, Nucl. Instr. Meth., A936 681 (2019)

    \bibitem{mupix8_2019}
        H.~Augustin et~al. {\em ``Performance of the large scale HV-CMOS pixel sensor MuPix8''}, arXiv:1905.09309, 2019

    \bibitem{MaNoehte}
        L. O. S. Noehte,
        \textit{``Microstrip Impedance Control in High Density Interconnects for the Mu3e Electrical Readout Chain''},
        Physikalisches Institut Heidelberg,
        2019,
        Available online at \texttt{https://www.psi.ch/de/mu3e/theses}.

    \bibitem{MaKroeger}
        Jens Kr{\"o}ger,
        \textit{``Readout Hardware for the MuPix8 Pixel Sensor Prototype and a Firmware-based MuPix8 Emulator''},
        Physikalisches Institut Heidelberg,
        2017,
        Available online at \texttt{https://www.psi.ch/de/mu3e/theses}.
            
    \bibitem{BaNoehte}
        L. O. S. Noehte,
        \textit{``Flexprint design and characterization for the Mu3e experiment''},
        Physikalisches Institut Heidelberg,
        2016,
        Available online at \texttt{https://www.psi.ch/de/mu3e/theses}.
	

    \bibitem{Hammerstad1} E. Hammerstad and O. Jensen, \textit{``Accurate Models for Microstrip Computer-Aided Design''}, 1980 IEEE MTT-S International Microwave symposium Digest, Washington, DC, USA, 1980, pp. 407-409.
% https://ieeexplore.ieee.org/document/1124303
%	H4ammerstad, E and Jansen, O,
%	\textit{Accurate Models for Microstrip Computer-Aided Design},
%	[p.~407-409],
%	1980.

    \bibitem{Hammerstad2} E. Hammerstad, \textit{``Computer-Aided Design of Microstrip Couplers with Accurate Discontinuity Models''}, 1981 IEEE MTT-S International Microwave Symposium Digest, Los Angeles, CA, USA, 1981, pp. 54-56.
% https://ieeexplore.ieee.org/document/1129818
%	Hammerstad, E,
%	\textit{Computer-Aided Design of Microstrip Couplers with Accurate Discontinuity Models},
%	[p.~54-56],
%	1981.
	
    \bibitem{atlc2}
	Dr. David Kirkby, G8WRB,
	\textit{``Arbitrary Transmission Line Calculator 2''},
	\texttt{http://www.hdtvprimer.com/KQ6QV/atlc2.html}

    \bibitem{MaDeflorin}
	M. Deflorin,
	\textit{``Helium cooling of Silicon Pixel Detector for Mu3e Experiment''},
	{Institut für Thermo- und Fluid-Engineering, Fachhochschule Nordwestschweiz},
	2019.
        Available online at \texttt{https://www.psi.ch/de/mu3e/theses}.



\end{thebibliography}
\end{document}